\documentstyle[aps,prb,twocolumn,epsf,rotate]{revtex}

\newcommand\be{\begin{equation}}
\newcommand\ee{\end{equation}}

\begin{document}
\twocolumn[\hsize\textwidth\columnwidth\hsize\csname
@twocolumnfalse\endcsname

\title{Interedge Phase Coherence in Quantum Hall Line Junctions}


\author{Aditi Mitra and S. M. Girvin}
\address{Department of Physics,
Indiana University, Bloomington, Indiana 47405-7105}
\date{\today}
\maketitle

\begin{abstract}

Kang et al.\ have recently observed a remarkable zero-bias anomaly in
the
spectrum for electron tunneling between two 2D electron gases separated
laterally by a narrow but high barrier in the presence of a
perpendicular quantising magnetic field. We argue that this is a result
of {\em interedge} phase coherence analogous to the {\em interlayer}
phase coherence seen in quantum Hall (QHE) bilayer systems. The
disruption of the QHE by the barrier is `healed' by strong coulomb
exchange enhancement of the weak tunneling through the barrier.

\end{abstract}


\vskip2pc]

\begin{figure}
\epsfxsize=2.0in \centerline{\epsffile{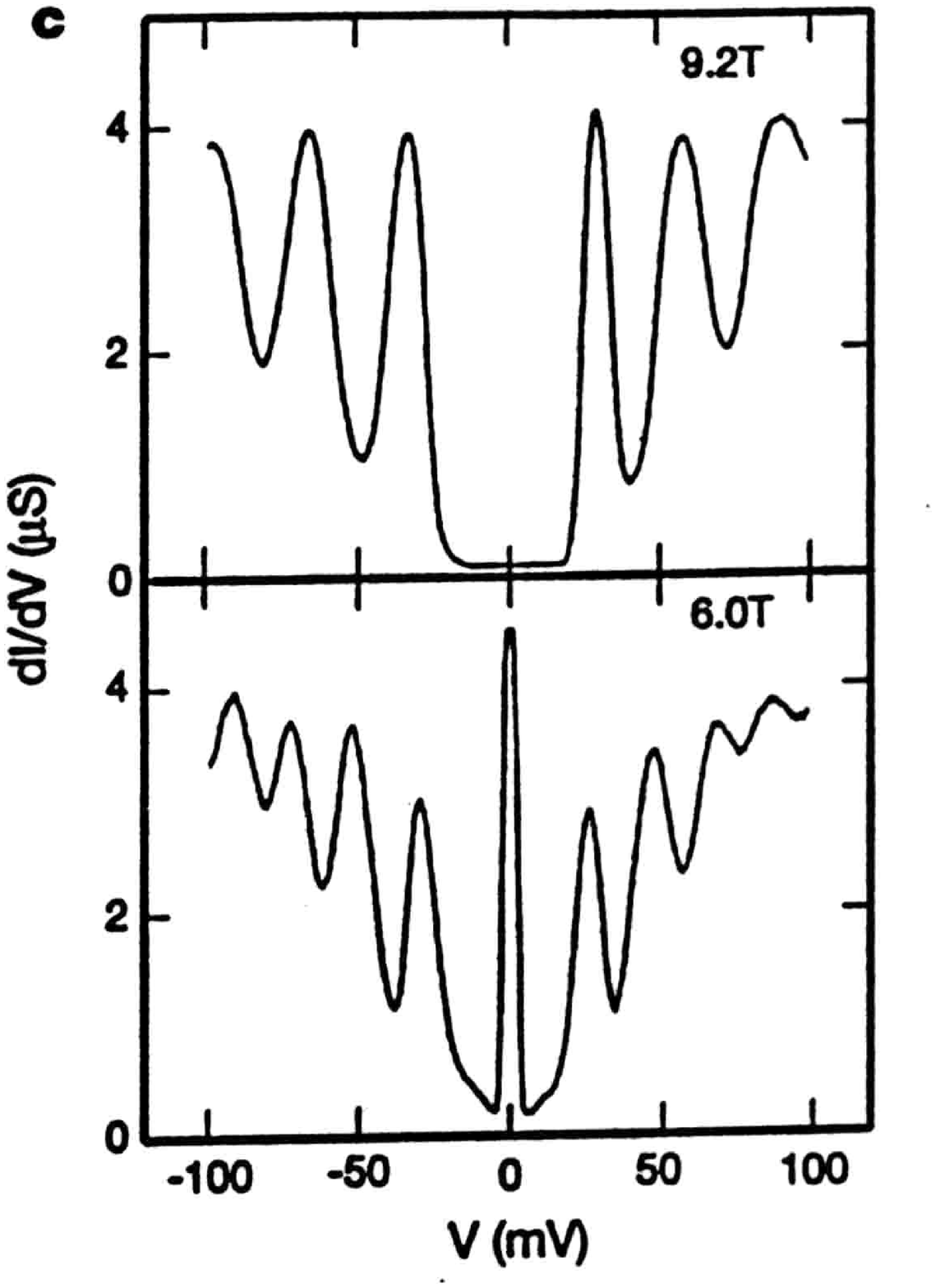}}
\caption{Experimental
tunnel conductance curves on and off the resonance filling factor
obtained by Kang {\it et al}. [After Ref.~[1] and W. Kang, private
communication.]
}
\label{fig:data}
\end{figure}

\begin{figure}
\epsfxsize=2.0in \centerline{\epsffile{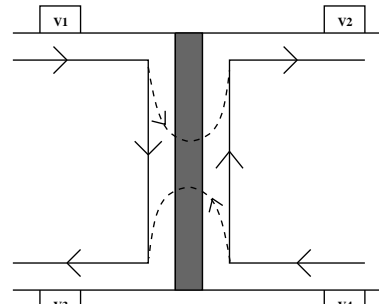}}
\caption{Top view of line junction.  The edge modes of the Hall bar are
assumed to be adiabatically connected to the corresponding barrier
modes.
}
\label{fig:linejunction}
\end{figure}

Kang et al.\ \cite{Kang} have recently observed a remarkable zero-bias
peak of height $G\sim 0.1 e^2/h$ in the differential tunneling
conductance between separate two-dimensional electron gases (2DEGs) for
certain values of magnetic field in the integer quantum Hall (QHE)
regime (see Fig.~(\ref{fig:data})). They used cleaved-edge overgrowth
to
construct a new sample geometry in which two adjacent 2DEGs in GaAs
quantum wells lying the in same plane are separated horizontally by a
narrow ($88\AA$) Al$_x$Ga$_{1-x}$As barrier forming a `line junction'
as
illustrated in Fig.~(\ref{fig:linejunction}).
\cite{Ho,Kane_Fisher,Ploog} In previous experiments the barrier was
defined electrostatically using top gates \cite{Haug} or narrow wires
\cite{Dynes} and so was much too thick ($0.1 - 10\, \mu$m) to allow
coherent tunneling.

A cleaved edge barrier was used by Chang and
collaborators \cite{Chang} to tunnel from a metallic
electrode into a quantum Hall edge
state. For the case of integer Landau level filling factor the
differential conductance is flat and independent of voltage at small
bias (as in a Fermi liquid), while for fractional filling factor
$\nu=1/3$, chiral Luttinger liquid physics yields an orthogonality
catastrophe and hence a differential conductance which {\em vanishes}
as
a power law in bias voltage. The resonance-like peak observed near zero
bias by Kang et al. for tunneling between two 2DEGs is quite different
from either of these two behaviors. A vastly sharper zero-bias anomaly
was discovered recently by Spielman et al. \cite{Eisenstein3} in
interlayer tunneling in a bilayer QHE system in which the 2DEGs are
separated {\em vertically} (i.e. in the MBE growth direction) rather
than horizontally by a Al$_x$Ga$_{1-x}$As barrier. This feature
is believed to be due to a novel broken symmetry producing
spontaneous interlayer phase coherence. \cite{Eisenstein3}

In this paper we
argue that the zero-bias anomaly seen by Kang et al.\ is due to the
analogous effect of {\em interedge} phase coherence. Because the
barrier
thickness is actually smaller than the spacing between electrons in the
plane, strong Coulomb correlations between the two sides can `heal' the
disruption in the QHE state caused by the barrier.  We develop a
bosonizaton scheme with parameters derived from Hartree-Fock
estimates to describe this process.

 Tunneling in the
absence of a magnetic field is a relatively simple matter. Electrons
approaching the barrier attempt to tunnel and either succeed or are
reflected and never return again to the barrier. The Landauer
conductance is $G_0 = 2\frac{e^2}{h}N_\perp |T|^2$ where the number of
transverse channels for the electron waves in a sample of width $W$ is
$N_\perp = k_FW/\pi$ and the transmission probability (averaged over
transverse channels) is $|T|^2$. Kang et al.\ found $1/G_0 \approx
450$k$\Omega$ for a sample with density $2\times 10^{11}$cm$^{-2}$ and
$W=100\mu$m. Using this and solving the Schr\"odinger equation for the
transmission we estimate the height of the $88\AA$ barrier to be $212$
meV which is in reasonable agreement with the barrier of $232$ meV
expected \cite{Davies} for Al$_x$Ga$_{1-x}$As if we use the average
value $x=0.3$ appropriate for the particular `digital' barrier in this
sample. \cite{Kang}

The physics of tunneling through a line junction in
the presence of a quantizing magnetic field is quite different.
\cite{Ho,Kane_Fisher} The magnetic field traps the electrons into
skipping orbits moving along the edge of the barrier in opposite
directions on the two sides, an effect previously observed in 3D
samples. \cite{Allen} Each electron thus makes not one, but many,
attempts to tunnel and can form a coherent superposition of states on
opposite sides. \cite{Ho} Kane and Fisher \cite{Kane_Fisher} have
modeled this situation assuming that the tunneling is a random function
of position along the line and showed that for integer filling
fractions, random tunneling is equivalent to random back scattering in
a
Fermi liquid with repulsive interactions and hence the states become
strongly localized. However because the barrier material in this
particular case is grown by MBE (during the first growth prior to
cleavage), disorder effects can be neglected in the barrier, at least
to a first approximation (see below however).
Strong justification for this assumption is
provided by vertical tunneling experiments (at $B=0$) which clearly
demonstrate nearly perfect momentum conservation for electrons passing
through such alloy barriers. \cite{Eisenstein1}

\begin{figure}
\epsfxsize=2.0in \centerline{\epsffile{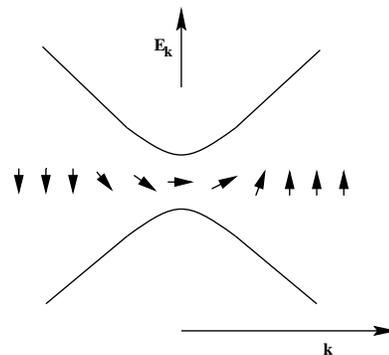}}
\caption{Dispersion of the edge modes in the vicinity of $k=0$.
Arrows indicate the orientation of the pseudospin vector for different
guiding center states labelled by wavevector $k$.
}
\label{fig:gap}
\end{figure}

To understand the electronic structure and particle transport in the
vicinity of the barrier we begin by assuming that the barrier is
infinitely long in the $y$ direction and presents a very high potential
$V_B(x)$ which is symmetric about $x=0$. Choosing the Landau gauge
${\bf
A} = eBx{\bf \hat y}/c$, the single particle wave functions are of the
form $\frac{1}{\sqrt{L_y}}e^{iky}\phi_{k}(x)$ where $\phi_{k}(x)$ is an
eigenfunction of the following Hamiltonian
\begin{equation}
H_k(x)= -\frac{\hbar^2}{2m}\frac{\partial^2}{\partial x^2} +
\frac{1}{2}m\omega_c^2(x-kl^2)^2 + V_{B}(x)
\label{eqn1}
\end{equation}
For each value of momentum $k$, there is a solution on the left side of
the barrier and another on the right side of the barrier. For the
special value of $k=0$, $H_k$ is reflection symmetric in $x$ and the
two
solutions will be degenerate in energy. If the barrier height is large
but finite, then these solutions will mix weakly and a gap $\Delta_0$
will open in the spectrum in the vicinity of the degeneracy point as
illustrated in Fig.~(\ref{fig:gap}). Away from the degeneracy point it
is not possible to tunnel through the barrier because states at the
same
momentum have very different energies.

In QHE bilayer systems it has proven extremely useful to introduce a
pseudospin 1/2 to represent the layer index degree of freedom.
The analog here represents the edge mode label. The
low-energy physics of the edge modes in this language is captured (for
the non-interacting case) by the 1D Hamiltonian
\be
H = \left(\begin{array}{cc}
-i\hbar v_0\partial_y&t_0\\
t_0&+i\hbar v_0\partial_y
\end{array}\right)
\label{eq:2by2H}
\ee
where $v_0 \sim 1.3\omega_c\ell$ is the bare edge mode velocity, $\ell$
is
the magnetic length, $\omega_c$ is the cyclotron frequency and for an
$88\AA$
barrier of height $212$ meV, the bare gap is $\Delta_0\equiv 2t_0 =
0.52$K. The ground state pseudospin orientation as a function of
momentum $k$ determined from the eigenfunctions of $H$ are illustrated
in Fig.~(\ref{fig:gap}). In the middle of a gap $\Delta$, the decay
length for the evanescent wave $\xi = 2\hbar v_0/\Delta$ which is on
the
scale of $1\mu$m for the bare gap and therefore much smaller than the
$\sim 100\mu$m barrier length in the Kang et al.\ samples.  We assume a
picture in which the transport current is carried by chiral edge modes
which
feed into the barrier modes.  If the
chemical potential lies within the gap for the barrier modes,
the edge modes of the Hall bar
fail to propagate along the barrier and must be perfectly
transmitted along the edge of the Hall bar (since the mode is chiral).
On the other hand, if the chemical potential lies
outside the gap, the edge mode connects smoothly onto the barrier
mode and is perfectly transmitted along the barrier
to the chiral edge mode on the opposite edge of
the sample as illustrated in Fig.(\ref{fig:linejunction}).

This picture, which assumes a QHE plateau and
$\sigma_{xx}=0$ in the bulk, predicts a
Landauer conductance peak of $e^2/h$. The observed peak
value of $0.1 e^2/h$ represents an enhancement of about a factor
of 2 over the $B=0$ conductance but is considerably smaller than the
ideal Landauer result. This may be a consequence of bulk transport
which
puts the system in an intermediate regime between the $B=0$ limit and
the idealized edge state picture of transport and allows electrons to
leak away from the barrier before they have had many opportunities to
tunnel back and forth to establish the coherent resonance gap.
Unfortunately the contact geometry in the experiment\cite{Kang}
did not permit a
determination $\sigma_{xx}$ in the QHE regime.  We also note in passing
that the conductance reported by Kang et al. used the voltage drop
between points  V1  and V2 shown in Fig.(\ref{fig:linejunction})
instead of between points V1 and V4.  The distinction is not
crucial when the peak height is much less than $e^2/h$ however.
As we discuss further below, inhomogeneities in the donor density
provide another possible explanation of the reduced height and large
width of the conductance peak.

A second peculiarity of the experimental data \cite{Kang} is that the
zero-bias peak occurs over a rather significant range
$\delta\nu \sim 0.3$ centered
at  filling factor $\nu^* \approx 1.35$.  The width $\delta\nu$ is
 well more than an order of magnitude wider than the expected value
$\delta\nu\sim \Delta_0/\hbar\omega_c$ given  the tiny bare gap of
$0.52$K. Takagachi and Ploog \cite{Ploog} (TP) have modeled the barrier
transport using a non-interacting electron tight binding model with
lattice constant $a=\ell/4$ and a near-neighbor hopping strength of
$t=\frac{\hbar^2}{2ma^2} =8\hbar\omega_c$. In a purely phenomenological
fit to the data, the barrier is modeled simply as a line of bonds with
reduced strength $t' = 0.02 t\sim 20$K. This is two orders of magnitude
larger than the bare tunnel amplitude $t_0=0.26$K computed from first
principles above. The tight-binding model used by TP
puts the level crossing slightly below the second Landau level, however
for non-interacting electrons with a realistic treatment of the
barrier, we find the crossing above the second Landau level.

In an attempt to resolve these paradoxes we have investigated the role
of strong Coulomb interactions along and across the barrier.  We find
in a
self-consistent Hartree calculation that the uncompensated background
charge underneath the barrier lowers the level crossing to below the
second
Landau level which is roughly
consistent with the experimental value of $\nu^*$
(assuming, as TP do,  complete spin polarization).
Using the self-consistent Hartree orbitals we further find that
Coulomb exchange leads to a large enhancement of the gap
in the Hartree-Fock (HF) spectrum to $\sim 10$K
(which is roughly consistent with the observed peak width).
However by bosonization (mapping the system onto a Luttinger liquid)
we find that the true gap is reduced by quantum fluctuations.

The HF variational ansatz for the many-body wave
function is
\begin{equation}
|\Psi\rangle = \prod_{k} (\cos{\frac{\theta_k}{2}}
c^{\dagger}_{k-\frac{\alpha}{2}\uparrow}+ \sin{\frac{\theta_k}{2}}
c^{\dagger}_{k+\frac{\alpha}{2}\downarrow})|0\rangle
\end{equation}
The order-parameter representing inter-edge phase coherence in this
state
\begin{equation}
\langle\Psi^{\dagger}_{\uparrow}(y)\Psi_{\downarrow}(y)\rangle =
e^{i\alpha y}\frac{1}{4\pi}\int dk\, \sin\theta_k
\end{equation}
tumbles at rate $\alpha$ along the $y$ direction.

We define a dimensionless variable $\varphi(y)$ that represents the
phase of the order-parameter as it tumbles. Evaluating the expectation
value of the number density operator in the HF state yields
\begin{equation}
\delta\rho(y) = \frac{1}{2\pi}\alpha =
\frac{1}{2\pi}\partial_y\varphi
\label{eq:numberdensity}
\end{equation}
which is the usual Luttinger liquid result.

Computing the expectation values of the kinetic, tunneling, Hartree,
and
exchange energies for the HF ansatz states yields the following
effective Hamiltonian density in terms of the order parameter field
$\varphi(y)$ and  the charge density imbalance $m_z \equiv
n_\uparrow(y) -
n_\downarrow(y)$
\begin{equation}
H_{\rm eff} = \frac{\rho_s}{2}\left(\partial_y \varphi
\right)^2 + \frac{\Gamma}{2} m_z^2 - t_0 \, \cos{\varphi(y)}
\end{equation}
We quantize this by making use of the
pseudospin commutation relations $[S_y,S_z] = i\hbar S_x$, which in
terms of the operators $\varphi$ and $m_z$, may be written as (assuming
$\varphi$ is small)
$
[\varphi,m_z/2] = i\hbar
$
Thus, $\varphi$ and $m_z/2$
are canonically conjugate to each other and the lagrangian density may
be written as
\begin{equation}
L = \frac{1}{8\pi g}[\frac{1}{c}\left
(\partial_t \varphi\right)^2
- c\left(\partial_y \varphi\right)^2] - t_0 \cos{\varphi(y)}
\label{eq:lagrange}
\end{equation}
In the absence of the tunneling term,
this is the lagrangian density of a Luttinger liquid with interaction
parameters $g = \sqrt{{\Gamma}/{\rho_s}}/2\pi$ and
collective mode velocity $c = 2 \sqrt{\rho_s \Gamma}$.
In this language,
the tunneling term is equivalent to $2k_{\rm F}$ backscattering.
An important property of our Luttinger liquid action
is the anisotropy of the coulomb interactions in the pseudospin
space which destroys the galilean invariance in the problem.

Our hartree-fock theory enables us to explicitly calculate the
parameters $\rho_s$ and $\Gamma$ and hence the Luttinger liquid
parameter g. The calculations have been done for a long range coloumb
potential, so the parameters $\rho$ and $\Gamma$ are
momentum dependent, with a $\ln{k}$ singularity in $\rho_s$, and under
these conditions the resulting action is not strictly a Luttinger
liquid.
To regulate this weak divergence we assume
metallic screening from nearby gates (or from the finite density of
states in the bulk, although this would violate our assumption of
$\sigma_{xx} = 0$)
and in that limit we have a Luttinger liquid whose parameters depend on
the Thomas-Fermi screening length. Table I
 lists the $k\rightarrow 0$
values for the Luttinger liquid parameter for different Thomas-Fermi
screening lengths. These calculations have been done for the
experimental barrier width of $88 \AA$ and also for a narrower barrier
width of $52 \AA$, the latter having a non-interacting tunnel splitting
which is almost 10 times that of the former.

In order to estimate $t_0\ell$ we choose the configuration of
the pseudo-spin corresponding to $\varphi=0$ everywhere and evaluate
the total tunneling energy in the corresponding HF state
$
t_0 \ell = \frac{1}{4\pi}\int dk\ell\, \Delta_k \, \sin{\theta_k}
$
where $\Delta_k/2$ is a measure of the amount by which the
energy is lowered as a result of tunneling. At $k=0$, $\Delta_0$ is
the bare tunnel gap.
The variational parameter $\sin{\theta_k}$ is unity at $k=0$, but falls
very rapidly to zero (within $|kl|< 0.1$) because the high
velocity in the region of the barrier makes it energetically
costly to have a coherent superposition of left and right orbitals for
large k.  Thus we can safely make the approximation, $t_0 \ell =
\frac{1}{4\pi} \Delta_0 \int dk\ell\, \sin{\theta_k} = 0.0053\,
\Delta_0 $.

Our HF calculations show an exchange enhanced gap of $\sim 10$K,
some $\sim 20$ times
larger than the bare gap. While this indicates the importance of
inter-edge coulomb interactions, the HF gap represents the excitation
gap
when the order parameter field is held fixed.  The true low-energy
charge excitations come from fluctuations of the order-parameter field.
 In
the absence of tunneling, the resulting U(1) symmetry of
Eq.(\ref{eq:lagrange})
guarantees that the charge excitations are gapless even though the
HF gap survives the limit of zero tunneling.
We see from Eq.(\ref{eq:numberdensity})
and Eq.(\ref{eq:lagrange}) that in the presence of tunneling, the
charged excitations are solitons having non-trivial topological charge
$\varphi(+\infty)-\varphi(-\infty)=\pm 2\pi$.

The classical expression for soliton energy (which becomes
exact in the strong interaction limit $g\rightarrow 0$)
is
$
\Delta_{\rm c} = 16\sqrt{t_0 \rho_s}
$
The sine-gordon model in Eq.(\ref{eq:lagrange})
is integrable and the expression for the soliton mass including
all quantum corrections is exactly known \cite{Zamolodchikov}
in the field theory limit $t_0/\hbar c\Lambda^2 \ll 1$ where
$\Lambda$ is the momentum cutoff.
The momentum cut off for the sine-gordon theory
that naturally emerges from our HF calculations is $\Lambda =
\frac{1}{2} \int dk \sin{\theta_k}$.   Physically, this corresponds to
the cutoff being proportional to the HF excitation gap, i.e.,
determined non-perturbatively by the interactions themselves.
Mathematically, we arrive at this expression
by comparing the coefficient of the $\cos{\varphi(y)}$ term in the
lagrangian obtained from our Hartree-Fock analysis, and that obtained
from bosonising the tunneling operator.   This procedure is not exact
but can be justified in the limit of small $g$.

The charge gap calculated for two different barrier widths is shown in
Table I.
For the barrier of width $88 \AA$ we find that
this number is about 1.3K. This is smaller than
the classical value of $4.8$K
but more than twice the size of the bare non-interacting value.
It is much too small to explain
the large width $\delta\nu$ for the zero-bias conductance peak seen in
the experiment.

These results imply that it is necessary to invoke
disorder.  We imagine that the MBE barrier is still smooth and
momentum conserving,  but note that the random variations in the donor
density which broaden the Landau levels in the bulk would lead to a
random
potential which would enter the Luttinger liquid action as {\em
forward}
scattering terms of the form $V_+(y) \partial_y \varphi + V_-(y) m_z$.
While this model ultimately flows under renormalization
onto the Kane and Fisher model \cite{Kane_Fisher}
of random tunneling along the barrier, we note that, paradoxically,
 the initial effect of this type of
 disorder is to {\em enhance} rather than destroy
the propogation along the barrier since even if the chemical potential
is
centered on the nominal charge gap, back scattering is effective only
in small regions where the random potential happens to pass through
zero.
This strong enhancement of the localization length could simultaneously
explain both the reduced height and the large observed width of
the zero-bias peak.  Some back scattering will occur over a range of
chemical potential determined by the strength of the random potential
rather than the nominal size of the charge gap.

Our results suggest that a second generation of samples with lower
disorder
and a somewhat thinner barrier could show the idealized behavior
modelled
here.  In such samples dynamical effects associated with the interedge
coherence should be observable.  While the dephasing time may be too
short
to observe the AC oscillations suggested by Ho \cite{Ho}, microwave
resonance absorption similar to that already seen \cite{Allen} in 3D is
likely to be visible.

We acknowledge numerous helpful discussions and suggestions from
L. Balents, W. Kang, H. Stormer, and especially S. Sachdev who
brought Ref. [\onlinecite{Zamolodchikov}] to our attention.
This work is supported by
NSF DMR-0087133 and the ITP at UCSB under NSF PHY99-07949.

\newpage

\mediumtext
\begin{table*}
\begin{tabular}{|c|c|c|c|c|c|c|c|c|c|c|} 
$w_b(\ell)$&$\Delta_0(K)$&$ q_{\rm TF}\ell $&$\frac{\Gamma^0}
{\ell}(\frac{e^2}{\epsilon \ell})$&$\frac{\Gamma^{\rm int}}{\ell}
(\frac{e^2}{\epsilon \ell})$&$\frac{\rho_s^0}{\ell}(\frac{e^2}
{\epsilon \ell})$&$\frac{\rho_s^{\rm int}}{\ell}(\frac{e^2}{\epsilon
\ell})$
& g&$\Lambda \ell$&$\frac{\hbar c}{\ell}$($\frac{e^2}{\epsilon
\ell}$)&$\Delta_{\rm c}(K)$  \\ \hline
0.84 & 0.52 & 0.05 & 4.21&1.53&0.107 &0.167 &0.73 & 0.033 & 2.51 & 1.26
\\
\hline
0.84 & 0.52 & 0.1  & 4.21&1.40&0.107 &0.108 &0.81 & 0.033 & 2.20 & 0.95
\\
\hline
0.50 & 4.00 & 0.05 & 3.77&1.16&0.095 &0.135 &0.74 & 0.049 & 2.13 & 6.6
\\
\hline
0.50 & 4.00 & 0.1  & 3.77&1.05&0.095 &0.089 &0.81 & 0.049 & 1.89 & 5.6
\\
\end{tabular}
\label{table:tableI}
\caption{Numerical results for model parameters for two values of
Thomas-Fermi screening constant $q_{\rm TF}$ and barrier thickness
$w_b$.
Calculations were done for a magnetic field of $B=6$T, magnetic length
$\ell=105\AA$ and cyclotron energy $\hbar\omega_{\rm c} = 10.4$meV.
$\Delta_0$ and $\Delta_{\rm c}$ are the bare and renormalized charge
gaps. $\Gamma_0$ and $\Gamma_{\rm int}$ are the bare and interaction
contributions to the Luttinger liquid stiffness parameter $\Gamma$.
$\rho_s^0$ and $\rho_s^{\rm int}$ are the bare and interaction
contributions to the Luttinger liquid compressibility parameter
$\rho_s$.
The parameter $g$ is the Luttinger exponent, $\Lambda$ is the estimated
cuttoff wavevector, and $c$ is the collective mode velocity in the
absence
of tunneling.
}
\end{table*}

\cleardoublepage
\narrowtext


\end{document}